\documentclass[12pt, a4paper]{article}

\usepackage{amsmath,amssymb,amsfonts}
\usepackage{graphicx}

\begin{document}

\title{Magnetogenesis in non-local models during inflation}
\author{E.V.~Gorbar${}^{1,2}$, T.V.~Gorkavenko${}^1$, V.M.~Gorkavenko${}^1$, O.M.~Teslyk${}^1$\\
${}^1$ \it \small Faculty of Physics, Taras Shevchenko National University of Kyiv,\\
\it \small 64, Volodymyrs'ka str., Kyiv 01601, Ukraine\\
${}^2$ \it \small Bogolyubov Institute for Theoretical Physics, National Academy of Sciences of Ukraine,\\
\it \small 14-b Metrolohichna str., Kyiv 03143, Ukraine}
\date{}

\maketitle
\setcounter{equation}{0}
\setcounter{page}{1}%

\begin{abstract}
The generation of magnetic fields during inflation in an electromagnetic model with a non-local form factor in Maxwell`s action is studied. The equations of
motion for the electromagnetic field are derived and solved. It is found that the conformal symmetry breaking due to the non-local form factor does not
lead to the generation of magnetic fields during inflation in the absence of interaction with the inflaton field. If such a coupling takes place, then
the presence of the form factor inhibits the generation of primordial magnetic fields compared to the case where the non-local form factor is absent.

\end{abstract}

\section{Introduction}

The quest for quantum gravity is one of the driving forces behind research in modern fundamental physics. It is well known that general relativity is a
non-renormalizable theory. Non-local theories of gravity provide an attractive possibility to regularize UV divergences and formulate a consistent theory. It is 
commonly believed that such a theory should resolve the singularities of black hole solutions in the general theory of relativity and shed light on the beginning and 
initial conditions of the Big Bang \cite{Kuz`min,Modesto,Tomboulis,Koshelev,Capozziello}.

The need for non-local theories to regularize high-energy divergences arises from the unitarity problem encountered in regularizations with a finite number of higher 
order derivatives. Indeed, although the addition of higher derivatives to the quantum field action regularizes UV divergences, these derivatives produce ghost states 
connected with the Ostrogradski instability \cite{Ostrogradsky} endangering unitarity. Non-local form factors with entire functions of the d`Alembertian
$\Box=g^{\mu\nu}\nabla_{\mu}\nabla_{\nu}$ avoid this problem because such models with infinite number of derivatives do not produce new poles in the propagators
of fields and, therefore, do not generate new physical degrees of freedom. It is noticeable also that vertices of the exponential form $e^{\Box}$, 
which is an entire function, appear in string field theory \cite{Witten}. The corresponding non-local gravity theories were studied in \cite{Biswas,Mazumdar}.

It is worth adding that non-locality in quantum field models appears not only as a means to regularize UV divergences but also in the derivation of
effective field theories accounting for the quantum corrections of heavy particles. The same applied to the quantum matter radiative corrections in semiclassical 
gravity (see, e.g., \cite{Barvinsky}). It was suggested in \cite{Woodard} that the vacuum polarization effects during inflation could be relevant for the generation
of cosmological-scale magnetic fields.

Recently magnetic fields with an extremely large coherence length measured in Megaparsecs were detected in cosmic voids through the gamma-ray observations of distant
blazars \cite{Neronov,Tavecchio,Taylor,Dermer,Caprini}. Such an extremely large coherence length suggests that these fields have a cosmological origin. Inflation can 
easily provide such a large coherence length for generated magnetic fields. However, the conformal symmetry of Maxwell`s action should be broken, otherwise, 
fluctuations of the electromagnetic
field cannot be enhanced in conformally flat Friedmann-Lema\^itre-Robertson-Walker (FLRW) background \cite{Parker}. In inflationary magnetogenesis studies, this breaking is 
usually taken in the form of the kinetic or axion coupling of the electromagnetic field with the inflaton field \cite{Turner,Ratra,Garretson,Dolgov}. Since non-local 
theories introduce an additional dimensional parameter, they necessarily break the conformal symmetry of Maxwell`s action too. Certainly, it would be very 
interesting if this breaking had been sufficient to generate a magnetic field of an appropriate strength during inflation without the need for the kinetic or axion 
coupling. This question provides the main motivation for the study in the present paper.

The paper is organized as follows. Magnetogenesis in a non-local electromagnetic model is considered in Sec.\ref{sec:non-local}. The 
obtained results are summarized in Sec.\ref{sec:conclusion}. Throughout the paper, we use the units with $\hbar\!=\!c\!=\!1$.

\section{Non-local electromagnetic  model}
\label{sec:non-local}

As we mentioned above, non-local models utilize form factors to ensure convergence at high momenta. In our analysis, we consider the exponential form factor
$e^{\Box/M^2}$, where $M$ is the regularizing mass parameter whose natural value is the Planck mass $M_{p}$. Then the corresponding Maxwell`s action 
takes the form
\begin{equation}
S=\int \sqrt{-g}\,d^4x 
\left[-\frac{1}{4}g^{\mu\alpha}g^{\nu\beta}F_{\mu\nu}e^{\Box/M^2}F_{\alpha\beta}+j_{\mu}A^{\mu}\right],
\label{Maxwell-action-regularized}
\end{equation}
where $g_{\mu\nu}$ is the spacetime metric, $F_{\mu\nu}=\nabla_{\mu}A_{\nu}-\nabla_{\nu}A_{\mu}$ is the strength tensor of the electromagnetic field $A_{\mu}$, and
$j_{\mu}$ is the electric current of charged matter fields.
Clearly, in view of the presence of the dimensional factor $M$ in the form factor, this regularized Maxwell`s action is not conformally symmetric which implies that 
electromagnetic fields, in principle, could be produced in an expanding FLRW background with scale factor $a(t)$, whose metric is given by $g_{\mu\nu}=\text{diag}(1,-a^2,-a^2,-a^2)$,  even in the absence of the interaction with charged matter fields.

Since $M$ is assumed to be larger than any other parameter in the model, including the Hubble constant $H$, the role of the conformal symmetry breaking due to the term
$e^{\Box/M^2}$ for magnetogenesis could be determined by approximating this non-local form factor with its first two terms in the Taylor expansion
$e^{\Box/M^2} \approx 1 + \Box/M^2$. Then we have
\begin{equation}
S=\int \sqrt{-g}\,d^4x \left[-\frac{1}{4}F_{\mu\nu}\left(1+\frac{g^{\sigma\rho}\nabla_{\sigma}\nabla_{\rho}}{M^2}\right)F^{\mu\nu}+j_{\mu}A^{\mu}\right]
\label{Maxwell-action-expansion}
\end{equation}
and obtain the following equations of motion for the electromagnetic field:
\begin{equation}
\nabla_{\mu}\left(1+\frac{g^{\sigma\rho}\nabla_{\sigma}\nabla_{\rho}}{M^2}\right)F^{\mu\nu}+j^{\nu}=0.
\label{equations-of-motion}
\end{equation}
Further, it is convenient to rewrite the above equation as follows:
\begin{equation}
\left(1+\frac{g^{\sigma\rho}\nabla_{\sigma}\nabla_{\rho}}{M^2}\right)\nabla_{\mu}F^{\mu\nu}
- \frac{1}{M^2}[g^{\sigma\rho}\nabla_{\sigma}\nabla_{\rho},\nabla_{\mu}]F^{\mu\nu}+j^{\nu}=0,
\label{equations-of-motion-1}
\end{equation}
where $[g^{\sigma\rho}\nabla_{\sigma}\nabla_{\rho},\nabla_{\mu}]$ is the commutator of the d`Alembertian and the covariant derivative.

Further,
\begin{equation}
-[\nabla_{\sigma}\nabla_{\rho},\nabla_{\mu}]=\nabla_{\sigma}[\nabla_{\mu},\nabla_{\rho}]+[\nabla_{\mu},\nabla_{\sigma}]\nabla_{\rho}
\label{commutators}
\end{equation}
and, for the commutator of covariant derivatives, we have
$$
[\nabla_{\mu},\nabla_{\sigma}]\phi_{\mu_1...\mu_k}=-\sum_{i=1}^k R^{\lambda}_{\,\,\,\,\mu_i\mu\sigma}\phi_{\mu_1...\mu_{i-1}\lambda\mu_{i+1}...\mu_k}.
$$
Since
$$
R_{\lambda\mu_i\mu\sigma}=H^2(g_{\lambda\mu}g_{\mu_i\sigma}-g_{\lambda\sigma}g_{\mu_i\mu})
$$
for de Sitter space \cite{Zee}, where $H$ is related to the Hubble constant in an inflationary expanding Universe, we find for the two commutators in Eq.(\ref{commutators})
\begin{align*}
 &   g^{\sigma\rho}[\nabla_{\mu},\nabla_{\sigma}]\nabla_{\rho}F^{\mu\nu}=-H^2\nabla_{\mu}F^{\mu\nu},\\
& g^{\sigma\rho}\nabla_{\sigma}[\nabla_{\mu},\nabla_{\rho}]F^{\mu\nu}=2H^2\nabla_{\mu}F^{\mu\nu}.
\end{align*}
Then Eq.(\ref{equations-of-motion-1}) takes the form
\begin{equation}
\left(1+\frac{g^{\sigma\rho}\nabla_{\sigma}\nabla_{\rho}}{M^2}\right)\nabla_{\mu}F^{\mu\nu}
+\frac{H^2}{M^2}\nabla_{\mu}F^{\mu\nu}+j^{\nu}=0
\label{equations-of-motion-2}
\end{equation}
or, equivalently,
\begin{equation}
\left(1+\frac{H^2}{M^2}\right)\nabla_{\mu}F^{\mu\nu}+\frac{\Box}{M^2}\nabla_{\mu}F^{\mu\nu}+j^{\nu}=0.
\label{equations-of-motion-3}
\end{equation}

Defining
\begin{equation}
\nabla_{\mu}F^{\mu\nu}=\frac{1}{\sqrt{-g}}\frac{\partial(\sqrt{-g}F^{\mu\nu})}{\partial x^{\mu}}=f^{\nu},
\label{function-f}
\end{equation}
Eq.(\ref{equations-of-motion-3}) implies the following equation for $f^{\nu}$:
\begin{equation}
\left(1+\frac{H^2}{M^2}\right)f^{\nu}+\frac{\Box}{M^2}f^{\nu}+j^{\nu}=0.
\label{equations-of-motion-4}
\end{equation}
Further,
\begin{multline}
\Box f^{\nu}=g^{\alpha\beta}\nabla_{\alpha}\nabla_{\beta}f^{\nu}=g^{\alpha\beta}\nabla_{\alpha}\left(\frac{\partial f^{\nu}}{\partial x^{\beta}}
+ \Gamma^{\nu}_{\rho\beta}f^{\rho}\right)=
\frac{1}{\sqrt{-g}}\partial_{\alpha}(\sqrt{-g}g^{\alpha\beta}\partial_{\beta}f^{\nu})
+\\ \frac{1}{\sqrt{-g}}\partial_{\alpha}(\sqrt{-g}g^{\alpha\beta}\Gamma^{\nu}_{\sigma\beta})f^{\sigma}+ 2\Gamma^{\nu}_{\sigma\alpha}g^{\alpha\beta}\partial_{\beta}f^{\sigma}
+g^{\alpha\beta}\Gamma^{\nu}_{\sigma\alpha}\Gamma^{\sigma}_{\rho\beta}f^{\rho},
\label{operator-f}
\end{multline}
where $\Gamma^{\nu}_{\alpha\beta}$ is the Christofell symbol
\begin{equation}
\Gamma_{\nu,\alpha\beta}=\frac{1}{2}\left(\frac{\partial g_{\nu\alpha}}{\partial x^{\beta}}+\frac{\partial g_{\nu\beta}}{\partial x^{\alpha}}
-\frac{\partial g_{\alpha\beta}}{\partial x^{\nu}}\right).
\label{Christofell-symbol}
\end{equation}

In the FLRW background and in the conformal time $\eta=\int^{t}dt'/a(t')$ the metric has the simple form $g_{\mu\nu}=a^2\eta_{\mu\nu}$, where $\eta_{\mu\nu}$ is
the Minkowski spacetime metric $\eta_{\mu\nu}=\text{diag}(1,-1,-1,-1)$. 
%In what follows, upper and lower indices are related via the Minkowski spacetime metric.
During inflation the scale factor is given by $a=-1/(H\eta)$ and the function $f^{\nu}$ defined in Eq.(\ref{function-f}) equals
$$
f^{\nu}=\frac{1}{\sqrt{-g}}\frac{\partial(\sqrt{-g}F^{\mu\nu})}{\partial x^{\mu}}
=\frac{\eta^{\nu\sigma}}{\sqrt{-g}}\eta^{\alpha\beta}\partial_{\alpha}\partial_{\beta}A_{\sigma},
$$
where 
$\eta_{\alpha\beta}$ is the Minkowski spacetime metric, 
$A^{\nu}=\eta^{\nu\sigma} A_\sigma$ is the electromagnetic field potential, and the 
Coulomb gauge
$A_0=0$ and $\text{div}\mathbf{A}=0$ was used. Thus, we have
\begin{equation*}
f^{\nu}=H^4\eta^4\eta^{\alpha\beta}\partial_{\alpha}\partial_{\beta}A^{\nu}
=H^4\eta^4\left\{\begin{array}{l} 0\hspace{7.4em} \text{for}\quad \nu=0\\
\eta^{\alpha\beta}\partial_{\alpha}\partial_{\beta}A^{i}\quad\quad\quad \text{for}\quad \nu = i,\quad i=1,2,3. \end{array} \right.  
\end{equation*}
By using Eqs.(\ref{operator-f}) and (\ref{Christofell-symbol}), one may check that $\Box f^{\nu}=0$ for $\nu=0$. Then, for vanishing current $j^{\nu}=0$, the equations 
of motion (\ref{equations-of-motion-4}) take the form
\begin{equation}
\left(1+\frac{H^2}{M^2}\right)f^{i}+\frac{\Box}{M^2}f^{i}=0,
\label{equations-of-motion-5}
\end{equation}
where $f^i=H^4\eta^4\eta^{\alpha\beta}\partial_{\alpha}\partial_{\beta}A^{i}$. Further, by using Eq.(\ref{Christofell-symbol}) and $g_{\mu\nu}=a^2\eta_{\mu\nu}$,
we find that Eq.(\ref{operator-f}) equals
$$
\Box f^i=\frac{1}{a^4}\left(\frac{(\partial^2_{\eta}a^2)}{2}+2(\partial_{\eta}a^2)\partial_{\eta}
+a^2\eta^{\mu\nu}\partial_{\mu}\partial_{\nu}\right)\frac{D^{i}}{a^4},
$$
where $D^i=\eta^{\alpha\beta}\partial_{\alpha}\partial_{\beta}A^i$. Then Eq.(\ref{equations-of-motion-5}) gives
\begin{equation}
D^{i}+\frac{H^2}{M^2}\left(4\eta\partial_{\eta}D^i+\eta^2\eta^{\alpha\beta}\partial_{\alpha}\partial_{\beta}D^i\right)=0.
\label{equations-of-motion-final}
\end{equation}

In the Coulomb gauge, only two transverse polarizations of the electromagnetic field remain. Then the electromagnetic vector-potential operator can be decomposed over 
the set of creation/annihilation operators as follows:
\begin{equation}
\label{quant-operator}
\hat{\mathbf{A}}(\eta,\mathbf{x})=\int\frac{d^{3}\mathbf{k}}{(2\pi)^{3/2}} \,\sum_{\lambda=\pm}\left\{ \boldsymbol{\epsilon}_{\lambda}(\mathbf{k})\hat{b}_{\lambda,\mathbf{k}}A_{\lambda}(\eta,\mathbf{k})e^{i\mathbf{k}\cdot\mathbf{x}}
+ \boldsymbol{\epsilon}^{*}_{\lambda}(\mathbf{k})\hat{b}^{\dagger}_{\lambda,\mathbf{k}}A^{*}_{\lambda}(\eta,\mathbf{k})e^{-i\mathbf{k}\cdot\mathbf{x}}\right\},
\end{equation}
where $\boldsymbol{\epsilon}_{\lambda}(\mathbf{k})$ is a set of two transverse circular polarization vectors, which satisfy the following conditions:
\begin{equation}
\mathbf{k}\cdot\boldsymbol{\epsilon}_{\lambda}(\mathbf{k})=0,\quad \boldsymbol{\epsilon}^{*}_{\lambda}(\mathbf{k})=\boldsymbol{\epsilon}_{-\lambda}(\mathbf{k}),
[i\mathbf{k}\times\boldsymbol{\epsilon}_{\lambda}(\mathbf{k})]=\lambda k \boldsymbol{\epsilon}_{\lambda}(\mathbf{k}).
\label{decomposition}
\end{equation}
The creation/annihilation operators satisfy the standard commutation relations
\begin{equation}
[\hat{b}_{\lambda,\mathbf{k}},\,\hat{b}^{\dagger}_{\lambda',\mathbf{k}'}]=\delta_{\lambda\lambda'}\delta^{(3)}(\mathbf{k}-\mathbf{k}').
\end{equation}
Substituting decomposition (\ref{quant-operator}) into Eq.~(\ref{equations-of-motion-final}) we obtain the equation governing the evolution of the mode function $A_{\lambda}$
\begin{equation}
(\partial^2_{\eta}+\mathbf{k}^2)A
+\frac{H^2}{M^2}\left[4\eta\partial_{\eta}+\eta^2(\partial^2_{\eta}+\mathbf{k}^2)\right](\partial^2_{\eta}+\mathbf{k}^2)A=0,
\label{equations-of-motion-ultimate}
\end{equation}
where, for simplicity, we suppress index $\lambda$ in the mode function. Making the change of variable $z=k\eta$, we obtain\vspace{-1em}
\begin{equation}
(\partial^2_{z}+1)A
+\frac{H^2}{M^2}\left[4z\partial_{z}+z^2(\partial^2_{z}+1)\right](\partial^2_{z}+1)A=0.
\label{equations-of-motion-ultimate-1}
\end{equation}

The form factor $e^{\Box/M^2}$ is an entire function of the d`Alembertian. This ensures that the photon propagator has only two poles, i.e., there are only two
electromagnetic modes at given momentum. Clearly, there are two solutions to Eq.(\ref{equations-of-motion-ultimate-1})
$$
A_{\pm}=C_{\pm}e^{\pm iz},
$$
which describe the usual free electromagnetic modes in the absence of any non-local form factor and conformal symmetry breaking in the free electromagnetic sector.
Any other solution to Eq.(\ref{equations-of-motion-ultimate-1}) is spurious and is related to the expansion of the form factor into a Taylor series and retaining only
its first two terms. Therefore, we conclude that the non-local form factor does not affect the free evolution of the electromagnetic field during inflation.
In other words, the form of Eq.\eqref{equations-of-motion-ultimate-1} with two operator-valued  multipliers $(\partial^2_{z}+1)$ acting on $A$  means that the inclusion of the non-local form factor
does not eliminate or modify the solutions for free  electromagnetic fields  in expanding FLRW Universe.

It is interesting to determine how the non-local form factor affects inflationary magnetogenesis in models where the electromagnetic field interacts with the inflaton
field $\varphi$. In the pseudoscalar inflation \cite{Garretson}, this interaction in Maxwell`s action (\ref{Maxwell-action-regularized}) is described by the current of the 
following form:
$$
j^{\nu}=\frac{I^{\prime}(\varphi)}{2\sqrt{-g}}\varepsilon^{\mu\nu\alpha\beta}F_{\alpha\beta}\partial_{\mu}\varphi,
$$
where $\varepsilon^{\mu\nu\alpha\beta}$ is the totally antisymmetric Levi-Civita tensor and $I(\varphi)$ is the coupling function of the electromagnetic field with the
inflaton field $\varphi$. Maxwell`s action with the non-local form factor (\ref{Maxwell-action-regularized}) implies the following equations of motion for the 
electromagnetic field:
\begin{equation}
\nabla_{\mu}e^{\Box/M^2}F^{\mu\nu}+j^{\nu}=0.
\label{equations-of-motion-inflaton}
\end{equation}
It is difficult to find explicit solutions to the above equation. However, qualitatively we could find out how the presence of the non-local form factor affects 
magnetogenesis. The form factor $e^{\Box/M^2}$ equals approximately the unity for eigenvalues of the d`Alembertian less than $M^2$ and rapidly increases for 
eigenvalues larger $M^2$. Therefore, to satisfy the above equation for given $j^{\nu}$, one would expect that $F^{\mu\nu}$ should be smaller in the case where the form 
factor $e^{\Box/M^2}$ is present. This means that the presence of the non-local form factor results in suppressed magnetogenesis in inflationary models.
\vspace{5mm}

\section{Conclusion}
\label{sec:conclusion}

The generation of magnetic fields in a non-local electromagnetic model with the form factor in Maxwell`s action in the form of the exponential of the d`Alembertian 
was studied during inflation. Solving the equations of motion for the electromagnetic field it was found that the conformal symmetry breaking induced by the non-local 
form factor does not lead to the generation of magnetic fields in an inflationary expanding Universe. 

Adding the interaction with the inflaton field   allows one to generate primodal magnetic fields.
Comparing   magnetic field generation in the models  of the pseudoscalar inflation without and with the non-local form factor shows that the presence of the form factor  inhibits the generation of primordial magnetic fields.
\vspace{5mm}

\centerline{\bf Acknowledgements}
\vspace{5mm}

The authors are grateful to O.O. Sobol and S.I. Vilchinskii for useful comments and discussions. The work of E.V.G. and O.M.T. was supported by the National Research Foundation of 
Ukraine Project No. 2020.02/0062.


\begin{thebibliography}{99}

\bibitem{Kuz`min} Yu.V.~Kuz`min, Sov. J. Nucl. Phys. {\bf 50}, 1011 (1989).

\bibitem{Modesto} L.~Modesto, Phys. Rev. D {\bf 86}, 044005 (2012).

\bibitem{Tomboulis} E.T.~Tomboulis, arXiv:hep-th/9702146.

\bibitem{Koshelev} A.S.~Koshelev, K.S.~Kumar, and A.A.~Starobinsky, Int. J. Mod. Phys. D {\bf 29}, 2043018 (2020).

\bibitem{Capozziello} S.~Capozziello and F.~Bajardi, Int. J. Mod. Phys. D \textbf{31},  2230009 (2022).

\bibitem{Ostrogradsky} M.~Ostrogradsky, Mem. Ac. St. Petersbourg {\bf VI}, 385 (1850). 

\bibitem{Witten} E.~Witten, Nucl. Phys. B {\bf 268}, 253 (1986).

\bibitem{Biswas} T.~Biswas, A.~Mazumdar, and W. Siegel, JCAP {\bf 0603}, 009 (2006).

\bibitem{Mazumdar} T.~Biswas, E.~Gerwick, T.~Koivisto, and A.~Mazumdar, Phys. Rev. Lett. {\bf 108}, 031101 (2012).

\bibitem{Barvinsky} A.~Barvinsky, Yu.~Gusev, G.~Vilkovisky, and V.~Zhytnikov, Nucl. Phys. B {\bf 439}, 561 (1995).

\bibitem{Woodard} T.~Propopec and R.P.~Woodard, Am. J. Phys. {\bf 72}, 60 (2004).

\bibitem{Neronov} A.~Neronov and I.~Vovk, Science {\bf 328}, 73 (2010).
		
\bibitem{Tavecchio} F.~Tavecchio, G.~Ghisellini, L.~Foschini, G.~Bonnoli, G.~Ghirlanda, and P.~Coppi, Mon. Not. R. Astron. Soc. {\bf 406}, L70 (2010).

\bibitem{Taylor} A.M.~Taylor, I.~Vovk, and A.~Neronov, Astron. Astrophys. {\bf 529}, A144 (2011).
		
\bibitem{Dermer} C.D.~Dermer, M.~Cavadini, S.~Razzaque, J.D.~Finke, J.~Chiang, and B.~Lott, Astrophys. J. Lett. {\bf 733}, L21 (2011).
		
\bibitem{Caprini} C.~Caprini and S.~Gabici, Phys. Rev. D {\bf 91}, 123514 (2015).

\bibitem{Parker} L. Parker, Phys. Rev. Lett. {\bf 21}, 562 (1968).

\bibitem{Turner} M.S.~Turner and L.M.~Widrow, Phys. Rev. D \textbf{37}, 2743 (1988).
		
\bibitem{Ratra} B.~Ratra, Astrophys. J. {\bf 391}, L1 (1992).
		
\bibitem{Garretson} W.~D.~Garretson, G.~B.~Field, and S.~M.~Carroll, Phys. Rev. D {\bf 46}, 5346 (1992).
		
\bibitem{Dolgov} A.D.~Dolgov, Phys. Rev. D \textbf{48}, 2499 (1993).

\bibitem{Zee} A. Zee, Einstein gravity in a nutshell, Princeton University Press, 889 pp. (2013).


\end{thebibliography}
\end{document}